\documentclass{rspublic}
\usepackage{graphicx, lscape}
\usepackage{amssymb, amsmath}

\usepackage{threeparttable}
\usepackage{natbib}
\usepackage{subfigure}

\begin{document}
\newcommand{\cc}{\mbox{cm$^{-3}$}}
\newcommand{\tauv}{\mbox{$\tau_V$}}
\newcommand{\ra}{\mbox{$\rightarrow$}}
\newcommand{\nhtwo}{\mbox{n$_{H_{2}}$}}

\def\spitz{{\em Spitzer}}               %  
\def\her{{\em Herschel}}               %   

\def\HI{H{\smc I}}
\def\HII{H{\smc II}}
%                                      molecules
\def\m17{M~17}               % M 17    
\def\cepa{Cepheus~A}               %  Ceph A    
\def\Htwo{H$_2$}               % H2    
\def\HtwoO{H$_2$O}             % H2O
\def\oHtwoO{ortho--H$_2$O}             % H2O 
\def\pHtwoO{para--H$_2$O}             % H2O 
\def\orthoHtwoO{o--H$_2$O}             % H2O 
\def\HtwoeiO{H$_2^{18}$O}             % H2(18)O 
\def\HtwoCO{H$_2$CO}           % H2CO 
\def\HtwoCS{H$_2$CS}           % H2CS 
\def\Hthreep{H$_3^+$}          % H3+
\def\HtwoDp{H$_2$D$^+$}        % H2D+
\def\DtwoHp{D$_2$H$^+$}        % D2H+
\def\Dthreep{D$_3^+$}        % D3+
\def\HthreeOp{H$_3$O$^+$}          % H3+
\def\HCOp{HCO$^+$}             % HCO+
\def\DCOp{DCO$^+$}             % DCO+
\def\HthCOp{H$^{13}$CO$^+$}    % H13CO+
\def\HtwCsiOp{H$^{12}$C$^{16}$O$^+$} % H12C16O+
\def\HCSp{HCS$^+$}             % HCS+
\def\HthCN{H$^{13}$CN}         % H13CN 
\def\HCfiN{HC$^{15}$N}         % HC15N 
\def\HtwCfoN{H$^{12}$C$^{14}$N}  % H12C14N 
\def\HNthC{HN$^{13}$C}         % HN13C
\def\HfoNtwC{H$^{14}$N$^{12}$C}  % H14N12C
\def\HCthreeN{HC$_3$N}         % HC3N 
\def\twCO{$^{12}$CO}           % 12CO
\def\thCO{$^{13}$CO}           % 13CO
\def\CseO{C$^{17}$O}           % C17O
\def\CeiO{C$^{18}$O}           % C18O
\def\twCsiO{$^{12}$C$^{16}$O}  % 12C16O
\def\thCsiO{$^{13}$C$^{16}$O}  % 13C16O
\def\twCeiO{$^{12}$C$^{18}$O}  % 12C18O
\def\thCeiO{$^{13}$C$^{18}$O}  % 13C18O
\def\CtfS{C$^{34}$S}           % C34S
\def\thCS{$^{13}$CS}           % 13CS 
\def\twCttS{$^{12}$C$^{32}$S}  % 12C32S
\def\tfSO{$^{34}$SO}           % 34SO
\def\ttSsiO{$^{32}$S$^{16}$O}  % 32S16O
\def\SOtwo{SO$_2$}             % SO2 
\def\tfSOtwo{$^{34}$SO$_2$}    % 34SO2
\def\SiO{SiO}             % SiO 
\def\Ntwo{N$_2$}               % N2
\def\Otwo{O$_2$}               % O2
\def\NtwoHp{N$_2$H$^+$}        % N2H+ 
\def\NHthree{NH$_{3}$}         % NH3
\def\CHthreeCCH{CH$_3$C$_{2}$H}     % CH3CCH
\def\CHthreeCN{CH$_3$CN}       % CH3CN
\def\CHthreeOH{CH$_3$OH}       % CH3OH
\def\CHfour{CH$_4$}       % CH4
\def\COtwo{CO$_2$}       % C02
\def\thCHthreeOH{$^{13}$CH$_3$OH}       % 13CH3OH
\def\twCHthsiOH{$^{12}$CH$_3$$^{16}$OH} % 12CH316OH
\def\CtwoH{C$_2$H}             % C2H
\def\CtwoS{C$_2$S}             % C2S
\def\CHp{CH$^{+}$}             % CH+
\def\Cp{C$^+$}             % C+
\def\Hp{H$^+$}             % H+
\def\Hep{He$^+$}             % He+
\def\CthreeHtwo{C$_3$H$_2$}    % C3H2
%                                      J transitions
\def\Jthoh{$J = 3/2 \to 1/2$}
\def\Johoh{$J = 1/2 \to 1/2$}
\def\Jtwel{$J = 12 \to 11$}
\def\Jelt{$J = 11 \to 10$}
\def\Jtn{$J = 10 \to 9$}
\def\Jne{$J = 9 \to 8$}
\def\Jes{$J = 8 \to 7$}
\def\Jss{$J = 7 \to 6$}
\def\Jsf{$J = 6 \to 5$}
\def\Jff{$J = 5 \to 4$}
\def\Jft{$J = 4 \to 3$}
\def\Jtt{$J = 3 \to 2$}
\def\Jto{$J = 2 \to 1$}
\def\Joz{$J = 1 \to 0$}
%                                      symbols
\def\WCO{W({\rm CO})}
\def\Wtw{W({\rm ^{12}CO})}
\def\Wth{W({\rm ^{13}CO})}
\def\dv{\Delta v}
\def\dvtw{\Delta v({\rm ^{12}CO})}
\def\dvth{\Delta v({\rm ^{13}CO})}
\def\NCO{N({\rm CO})}
\def\Nth{N({\rm ^{13}CO})}
\def\Ntw{N({\rm ^{12}CO})}
\def\NtwCsiO{N({\rm ^{12}C^{16}O})}
\def\NthCO{N({\rm ^{13}CO})}
\def\NthCsiO{N({\rm ^{13}C^{16}O})}
\def\NtwCeiO{N({\rm ^{12}C^{18}O})}
\def\intCO{\int T_R({\rm CO})dv}
\def\inttwCsiO{\int T_R({\rm ^{12}C^{16}O})dv}
\def\intthCsiO{\int T_R({\rm ^{13}C^{16}O})dv}
\def\inttwCeiO{\int T_R({\rm ^{12}C^{18}O})dv}
\def\NHtwo{N({\rm H_2})}
\def\Wtw{W_{12}}
\def\Wth{W_{13}}
\def\kappanu{\kappa_{\nu}}
\def\phinu{\varphi_{\nu}}
\def\taunu{\tau_{\nu}}
\def\dv{\Delta v}
\def\dvFWHM{\Delta v_{FWHM}}
\def\vLSR{v_{LSR}}
\def\Rsol{R_\odot}
\def\Msol{M_\odot}
\def\MMsol{\ts 10^6\ts M_\odot}
\def\MCO{M_{\rm CO}} 
\def\Mvir{M_{\rm vir}}
\def\TAstar{T^*_A}
\def\TAstartwCO{T^*_A(^{12}{\rm CO})}
\def\TAstarthCO{T^*_A(^{13}{\rm CO})}
\def\TAstarCeiO{T^*_A({\rm C}^{18}{\rm O})}
\def\TRstar{T^*_R}
\def\TexCO{T_{ex}({\rm CO})}
\def\Trms{T_{rms}}
%                                      			units
\def\d{^\circ}
\def\h{^{\rm h}}
\def\mi{^{\rm m}}
\def\s{^{\rm s}}
\def\mum{\ts \mu{\rm m}}
\def\mm{\ts {\rm mm}}
\def\cm{\ts {\rm cm}}
\def\percm{\ts {\rm cm}^{-1}}
\def\m{\ts {\rm m}}
\newcommand\kms{\rm{\, km \, s^{-1}}}
\def\K{\ts {\rm K}}
\def\Kkms{\ts {\rm K\ts km\ts s^{-1}}}
\def\kHz{\ts {\rm kHz}}
\def\MHz{\ts {\rm MHz}}
\def\GHz{\ts {\rm GHz}}
\def\pc{\ts {\rm pc}}
\def\kpc{\ts {\rm kpc}}
\def\Mpc{\ts {\rm Mpc}}
\def\cmsq{\ts {\rm cm^2}}
\def\pcsq{\ts {\rm pc^2}}
\def\dsq{\ts {\rm deg^2}}
\def\debye{\ts10^{-18}\ts {\rm esu}\ts {\rm cm}}
\def\swash2o{$1_{10} - 1_{01}$}             

%                                      			journals
%                                      			mathe
\let\ap=\approx
\let\ts=\thinspace

\title[Water and Star Formation]{Water in Star and Planet Forming Regions}
\author[E. A. Bergin, E. F. van Dishoeck]{Edwin A. Bergin$^{1}$ and Ewine F. van Dishoeck$^{2,3}$}
\affiliation{$^1$Department of Astronomy, University of Michigan, 500 Church St. , Ann Arbor MI 48197, USA\\
$^2$Leiden Observatory, Leiden University, PO Box 9513, 2300 RA Leiden, The Netherlands \\
$^3$Max Planck Institut f¬ur Extraterrestrische Physik, Giessenbachstrasse 1, 85748 Garching, Germany
}
\maketitle

\begin{abstract}
\noindent In this paper we discuss the astronomical search
for water vapor in order to understand the disposition of
water in all its phases throughout the process of star and
planet formation.   Our ability to detect and study water
vapor has recently received a tremendous boost with the
successful launch and operations of the {\em Herschel Space
Observatory}.  \her\ spectroscopic detections of numerous
transitions in a variety of astronomical objects, along with
previous work by other space-based observatories,
will be threaded throughout this contribution.  In particular,  we present 
observations of water tracing the earliest stage of star 
birth where it is predominantly frozen as ice.  When a star is born the 
local energy release by radiation liberates ices in
its surrounding envelope and powers energetic outflows
that appear to be water factories.  In these regions water plays
an important role in the gas physics.   Finally, we end
with an exploration of water in planet forming disks surrounding young stars.   The availability of  accurate molecular data (frequencies, collisional rate coefficients, and chemical reaction rates) are crucial to analyze the observations at each of these steps. 
\end{abstract}

%\section{}
%\subsection{}

\section{Introduction}
Because of its importance for life on Earth, water is
one of the most important molecules in the solar system and beyond.
However, the universe is not filled with water and thus the creation of H$_2$O from oxygen and hydrogen atoms, along with its disposition 
 within each stage of the formation of stars and planets is a question with
clear association to our own origins.      The search for greater understanding of this ``water cycle'' is complicated by the fact that water in the Earth's atmosphere impedes direct observation of water emissions in star-forming gas, except for some high excitation (masing) transitions\footnote{Water vapor has long been detected from the ground in the atmospheres of cool stars \citep{wsr64, sn65}.  However, the physical conditions of star forming clouds populate rotational transitions of colder water (T $<$ 800 K), which are generally blocked by the Earth's atmosphere.  }.   Thus we rely on space-based observatories to gain access to the full spectrum of water vapor.     Following water in star and planet formation involves a decrease in size scale of over 10$^5$ (from parsec to below an astronomical unit; 1 pc = 206265 AU = 3.086 $\times 10^{18}$ cm; one AU = distance from Earth to the Sun = 1.5 $\times 10^{13}$ cm).   The gas involved in the gravitational collapse undergoes sharp increases in physical properties spanning many orders of magnitude in density and several in temperature.     The water chemistry and its resulting emissions 
track these changes, leading water to be a true astrophysical probe, along with leaving a trail of the water cycle.

In this paper we will outline some of our basic understanding of the water cycle gleaned through astronomical observations over the past several decades.   
To a large extent  no single observatory is capable of capturing the entire picture and much of this understanding is built upon analyses performed by numerous individuals using observations of H$^{18}$O from ground-based observatories and H$_2$O (along with the isotopologues) from space-based platforms such as the {\em Infrared Space Observatory} (ISO),  the {\em Submillimeter Wave Astronomy Satellite} (SWAS), Odin, the {\em Spitzer Space Telescope}, and today with the {\em Herschel Space Observatory}.    The aim of this paper is to do our best to elucidate our basic understanding of water vapor in space.    However, for completeness, we also refer the reader to additional articles that summarize some of the relevant results from these missions \citep{hjalmarson03, evd04, melnick09, klaus10}.   The legacy of \her\ is still unfolding, but initial results are summarized in \citet{evd_wish}.
 
In \S~2 we briefly discuss star and planet formation, along with the observed gas physical conditions and dominant chemical processes.  \S~3 outlines the use of water vapor emission as a probe of astrophysics.  We will also discuss the disparate capabilities of the space-based observatories along with an outline of the chemistry of water in space.  
In \S~4 we discuss the observational constrains on water vapor and star formation with a relation to our understanding of its formation/destruction and primary phase.   Finally we summarize our high level understanding in \S~5.

\section{Star Formation and Properties of Molecular Gas}
\label{sec:sf}
Below we outline a general understanding of star and planet formation.   Table~1 lists some of the relevant physical properties associated with each stage outlined below.    The role and disposition of water will be discussed within this context.

\underline{\em Molecular Cloud:}
  Stars are born in molecular clouds with typical densities of a few thousand H$_2$ molecules per cubic centimeter, gas temperatures of $\sim 10 - 20$~K, with typical masses of $10^3 - 10^5$ M$_{\odot}$.   The clouds are predominantly gaseous in composition with a solid-state component labelled as dust grains that are silicate and carbonaceous (graphite, amorphous carbon) in composition \citep{draine03}.
  Molecular clouds exist over scales of tens of parsecs, but exhibit definite substructure with stars being born in denser (n $> 10^5$ \cc ) cores with typical sizes of 0.1 pc.     The boundaries of the parsec-sized cloud are generally defined by the extent of CO emission, which is the primary observable gas-phase molecular constituent at large scales.    
  
\underline{\em Pre-Stellar Cores:}
  Dense cores (0.1 pc) have been primarily characterized as clumps of concentrated sub-millimeter dust emission or infrared absorption \citep{andre_ppiv, alves_b68, bacmann_iso} and through transitions of simple molecules with high-dipole moments \citep[e.g. \NHthree , \NtwoHp , CS; ][]{bm89, cbmt02, bt_araa, difran_ppv}.   For the most part these molecules are formed via ion-molecule chemistry in the gaseous state (e.g. CS, \NtwoHp ).   However, prior to stellar birth, the condensation of the core increases the central density by 2--3 orders of magnitude, while still remaining quite cold (T $\sim$ 10~K).     Because  the density increase leads to more frequent collisions with cold dust grains (T$_d \sim$ 10~K), the central regions of the core are dominated by the freeze-out of atoms and molecules from the gas onto dust grains \citep[][and references therein]{bt_araa}., forming an ice mantle coating on the solid silicate grains \citep{jw84, hhl92}.  There is also  the possibility of subsequent reactions within the ice mantle that lead to greater complexity \citep{ta87, hhl92, gwh08, cr09}.  This is provided the dust temperature is below the relevant sublimation temperature for a given atom or molecule.   
    
     \underline{\em Protostars and Outflows:}
   Molecular cloud cores are rotating and, upon gravitational collapse, the infalling envelope flattens to a disk and a central source \citep[see ][for greater discussion]{als87, awb93}.    At this time the central source releases gravitational energy and heats its surrounding envelope.   The  central region is characterized by gas phase emissions at high temperature ($\ge 100$~K) and it is called the ``hot core''.  The hot core exhibits a diverse chemistry powered by evaporation of the ice mantle along with subsequent gas-phase reactions \citep{blake87, bcm88, hvd09}.    During the protostellar stage the forming star begins to eject material and the envelope is disrupted along the poles and the star begins the process of destroying its natal envelope.  This mass ejection powers what are known as bipolar molecular outflows that are characterized by two extremes (that likely span a continuous range of physical properties).   At one extreme is a more extended (sub-parsec to parsec scale) molecular outflow of entrained gas with temperatures somewhat elevated above the natal cloud (e.g. Table~1).   The other extreme are smaller regions of emission where faster moving material has impacted slower moving gas leading to a shock.  
Shocked gas is characterized by temperatures in excess of several hundred K; crucially this can power reactions with barriers that were inactive in the cold (T $\sim 10 - 20$~K) gas and the release of ice mantles and more refractory material into the gas.   

     \underline{\em Protoplanetary Disk:}
As the envelope dissipates, the gas-rich disk becomes exposed. The disk surface is directly irradiated by energetic UV and X-ray photons from the star, but also by the external (interstellar/local) radiation field.    Observational systems in this stage are often called T Tauri stars.  Disks have strong radial and vertical gradients in physical properties with most of the mass residing in the middle of the disk, labeled as the midplane \citep[e.g., ][]{bs90, calvet91, cg97, calvet_ppcd}.    It is the midplane that is the site of planet formation.   
 The density of the midplane significantly exceeds that of the dense natal core, by many orders of magnitude.
  The midplane is in general colder than the disk upper layers, which can also be called the disk atmosphere. Finally, the outer radii of the disk (r $>$ 10 AU) has reduced pressure and temperature, but contains most of the mass.    Molecular observations are only beginning to probe this stage but a general picture has emerged.  At large radii (r $>$10--30 AU) molecules are destroyed on the exposed disk surface due to photodissociation.  In the midplane the temperatures are below the sublimation temperature and most molecules are frozen as ices.  In between are warm molecule-rich (T $\sim$ 50 K) layers where some ices are able to evaporate \citep[e.g. CO; ][]{aikawa_vanz02, kamp05}.   Inside of 30 AU the midplane temperature progressively increases and a so-called ``snow-line'' is believed to exist.  Inside the snow-line a given species would exist in the gas and beyond as ice.  This line might be species specific.

\begin{table}
\label{tab:sf}
\begin{threeparttable}
\caption{Estimated Physical Properties of Star and Planet Forming Gas\tnote{1}}
\begin{tabular}{lllll}
\hline
\multicolumn{1}{c}{Stage} &
\multicolumn{1}{c}{Volume Density\tnote{2}} &
\multicolumn{1}{c}{T$_{gas}$} &
\multicolumn{1}{c}{$\Delta v$\tnote{3}} &
\multicolumn{1}{c}{Notes\tnote{4}} \\\hline\hline
Extended Cloud & $\sim 1000$ cm$^{-3}$ & 10--15~K & 1 - 3 km s$^{-1}$  &  \\
Pre-Stellar Cores & $10^5 - 10^6$ cm$^{-3}$ & 10--15~K & 0.3 - 1.5 km s$^{-1}$  & (1) \\
Protostar & &  &   &  \\ 
\hspace{0.5cm}Hot Core & $> 10^6$ cm$^{-3}$ & 100--300~K & 5 -- 15 km s$^{-1}$  & (2) \\
\hspace{0.5cm}Extended Outflow & $10^3 - 10^4$ cm$^{-3}$ & 10--100~K & 5 -- 15 km s$^{-1}$  & (3) \\
\hspace{0.5cm}Shocked Gas & $10^5 - 10^6$ cm$^{-3}$ & 100--1000~K & 10 -- 50 km s$^{-1}$  & (4) \\
Protoplanetary Disk & $10^6 - 10^{15}$ cm$^{-3}$ & 10--2000~K & 1--5 km/s  & (5)\tnote{5} \\
\hline
\end{tabular}
\begin{tablenotes}[para]
\footnotesize
\item[1] Parameters are provided to illustrate the range of values inferred from emission studies and
may not be representative of the extremes.\\
\item[2] Density as inferred from the excitation of high dipole moment molecules (similar to water vapor).\\
\item[3] Full width at half maximum of spectrally resolved molecular emission lines, usually representative of turbulent or dynamical motions.\\
\item[4] (1) \citet{bt_araa}; (2) \citet{beuther_ppv} and \citet{ceccarelli_ppv}; (3) Referring to the swept up molecular outflow with properties summarized by \citet{lada85} and \citet{bachiller96}; (4) \citet{gg99}, \citet{nisini02}, \citet{nisini07}; (5) \citet{nomura07}, \citet{oberg_discs}.\\
\item[5] If line broadening is dominated by Keplerian rotation, as generally assumed, then line widths should increase if the molecular emission originates from closer to the star.  Values reported are for unresolved (or barely resolved) disks.
\end{tablenotes}
\end{threeparttable}
\end{table}

\section{Molecular Astrophysics and Chemistry of Water}
\label{sec:ma}

\subsection{H$_2$O Emission and Observations}
Water is a highly asymmetric molecule with its energy levels labelled by a set of 3 quantum numbers, J$_{\rm K_A, K_C}$.
Water is comprised of two identical hydrogen nuclei, each with nuclear spin.  As such, due to Fermi-Dirac statistics, it exists in two forms or spin isotopomers.   One with total spin of unity is \oHtwoO , while the other with total spin of zero is \pHtwoO .   The energy levels are different for each with \oHtwoO\ having {\rm $K_A + K_C$ = odd}, and \pHtwoO\ characterized by  {\rm $K_A + K_C$ = odd}.
Transitions across ortho and para levels are strictly forbidden \citep{dennison27}, and the conversion amongst the spin isotopomers occurs via reactions which transfer an H atom.   In the high temperature limit the ratio of these two species is set by the ratios of the spin statistical weights resulting in an ortho-to-para ratio (OPR) of 3:1.   The ground state of \pHtwoO\  is 34.3 K below that of \oHtwoO .  Thus it is possible, if water forms at temperatures below this value that \pHtwoO\ can dominate.  Comets are known to have an OPR below 3 possibly coincident with formation at $\sim 30$~K.  However, the mechanisms for ortho-to-para conversion in the gas and on the grain surface are not well characterized \citep[see ][for a discussion]{lis_op}.

H$_2$O, as a light hydride, has large energy level spacings when compared to heavier molecules (e.g. CO) and its main transitions lie in the sub-millimeter to far-infrared wavelengths.     In addition, with a high dipole moment of 1.85 Debye and large line frequencies, it has rapid spontaneous decay coefficients.
This can lead to large line opacities for small columns.
To elucidate  this point, the lower state column ($N_l$) of \oHtwoO\ $1_{10} \rightarrow 1_{01}$ is related to the line center optical depth ($\tau_0$) by $N_l({\rm o-H_2O}) = 4.89 \times 10^{12}\tau_0\Delta v$ \citep{plume04}.   In this formula $\Delta v$ is the full-width-at-half-maximum of the line in km s$^{-1}$. $N_l$  represents the total \oHtwoO\ column provided the gas density does not exceed \nhtwo\ $> 10^{7}$ \cc (assuming $\Delta v = 1$ km s$^{-1}$).   Thus $\tau_0 = 1$ for a N(\oHtwoO ) $\sim 5 \times 10^{12}$ cm$^{-2}$.    To place this in perspective a typical cloud would have regions with a total \Htwo\ column of $10^{22}$ cm$^{-2}$.   Thus water emission is optically thick in a representative cloud (again assuming $\Delta v = 1$ km s$^{-1}$) even at very low abundances of n(\oHtwoO )/n(\Htwo ) $\sim 5 \times 10^{-10}$.  The solar elemental oxygen abundance is 9.8 $\times 10^{-4}$ \citep[relative to \Htwo ;][]{asplund09}, while the average oxygen abundance $\sim 1.15 \times 10^{-4}$ in a large sample of F, G, and K stars \citep{pm11}.  Thus, this amount of water traps only a tiny fraction of the available oxygen.

\begin{figure}
\centering
\includegraphics[width=\textwidth]{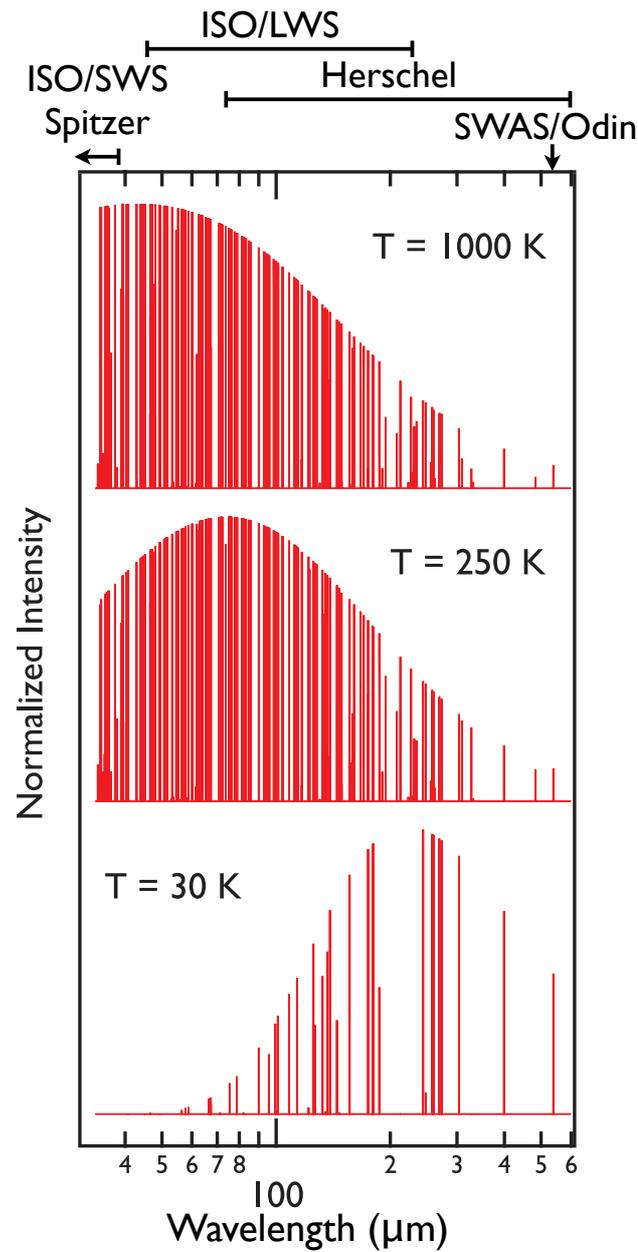}
\caption{Optically thin emission spectrum of water vapor in LTE at 30~K, 250~K, and 1000~K.   Intensities for each temperature are normalized to emphasize the range of spectral lines as a function of wavelength as opposed to the overall intensity. 
To create this model spectrum  we made use of the myXCLASS program
  (http://www.astro.uni-koeln.de/projects/schilke/XCLASS) which accesses
  the JPL
  \citep[][http://spec.jpl.nasa.gov]{pickett98} molecular data base for water vapor line assignment. }
\label{fig:h2ospec}
\end{figure}

Water has long been observed in star and planet forming regions via its masing emissions at centimeter wavelengths \citep{cheung69}.     In this paper we will focus on observations of thermal water vapor emission and in particular its rotational transitions which are attuned to the physical conditions of star forming regions.  
Fig.~\ref{fig:h2ospec} shows the optically thin LTE emission spectrum of water from 30 to 600 $\mu$m ($\sim 500$ GHz - 10 THz) for gas temperatures of 30, 250, and 1000 K.     First it is clear that the spectral shape, or emission envelope, exhibits a strong temperature dependence.   Lower temperature gas emits primarily in the sub-millimeter where the ground-state lines lie (557 GHz for \oHtwoO , and 1113 GHz for \pHtwoO ), while at T $\sim$ 250~K the emission peaks in the far-infrared.  Finally very hot gas has numerous emission lines that extend to mid-infrared wavelengths and beyond the right-hand edge of the plot.   This clearly demonstrates that water has numerous emission lines and the relative intensities bear information on the gas temperature.    Of course that is not the complete story as water emission will also depend on density and column density.   For water emission to truly reach LTE it requires very high densities (at 1000~K it will be $>$ 10$^{13}$ \cc ).    Thus this spectrum shows the maximum number of emissive lines for water over this spectral range and the true number of lines for typical conditions will be well below this level.

Another aspect is the relative utility of a given observatory to trace water vapor emission.  Essentially the different platforms are capable of probing a specific range of physical conditions and thus explore specific aspects of star formation physics (e.g. Table~\ref{tab:sf}).    The top of Fig.~\ref{fig:h2ospec} presents the spectral grasp of each of the respective space-based observatories.    \spitz\ with its wavelength coverage and moderate spectral resolution enabled the detection of very hot gas which is found in outflows \citep{melnick08} and also in the innermost regions of protoplanetary disks \citep{carrnajita08}.     ISO was best suited for the  detection of warm and hot water in shocks and also in the hot cores around massive stars \citep[][and references therein]{evd04, nisini05}.   In the case of \spitz\ and ISO the lines are generally not velocity resolved.   SWAS and Odin both were tuned to the ground state line of \oHtwoO\ which is $\sim 26.7$~ K above the ground state.  This line was sensitive to the cold (T $\sim 10-40$~K) and warm (T $\sim 100$~K) gas which can be present in outflows, hot cores, and quiescent gas \citep{hjalmarson03, melnick09}.   These observatories used heterodyne techniques allowing for enough spectral resolution to resolve the emission lines, even in the extended cloud.     However, both SWAS and Odin had moderate angular resolution and were not sensitive enough to detect water in pre-stellar cores or disks.
In addition, the analysis was hampered by observing only one emission line (supplemented by the ground state of ortho-H$_2^{18}$O).   Finally, \her\ has tremendous capability to explore all the components of star formation with the greatest angular resolution to date through instruments that offer both high and low spectral resolution; it also has receivers that have noise temperature only just above the quantum limit.  There is a dedicated Key Program on the topic of water:  ``Water in Star-forming Regions with the {\em Herschel} Space Observatory'' (WISH; E. van Dishoeck, PI).  A summary of the initial WISH results is given by \citet{evd_wish} and some results are further discussed below.

\begin{table}
\begin{threeparttable}
\caption{Main Space-Based Platforms for Observations of Rotationally Excited H$_2$O}
\begin{tabular}{lccrrl}
\hline
\multicolumn{1}{c}{Observatory\tnote[1]} &
\multicolumn{1}{c}{Spectral Range} &
\multicolumn{1}{c}{Angular Resolution} &
\multicolumn{2}{c}{R} &
\multicolumn{1}{c}{Ref.\tnote{2}} \\
\multicolumn{1}{c}{} &
\multicolumn{1}{c}{$(\mu m)$} &
\multicolumn{1}{c}{} &
\multicolumn{1}{c}{$(\lambda/\Delta \lambda$)} &
\multicolumn{1}{c}{(km/s)} &
\multicolumn{1}{c}{} \\\hline\hline
ISO & & & & \\
\hspace{0.1in} SWS & 2.4 -- 45.2   & $14''  \times 20''$ -- $20'' \times 33''$  & 1500 & 200 & (1) \\
\hspace{0.1in} SWS & 11.4 -- 44.5  & $10'' \times 39''$ -- $17'' \times 40''$  & 30000 & 10 & \\
SWAS & 538.6 & 3.3' $\times$ 4.5$'$ & $5 \times 10^5$ & 0.5 & (2)\\
Odin\tnote{3}   & 517 -- 617 & 2.1$'$ & $> 5 \times 10^{5}$ & 0.05 -- 0.5 & (3)\\
\spitz\ & & & &\\
\hspace{0.1in} IRS SH/LH & 10 -- 36 & $5'' - 10''$ & 600 & 5000 & (4) \\
\her\ & & & &\\
\hspace{0.1in} SPIRE & 194 -- 671  & 17$''$ - 41$''$ & 370 -- 1300 & 230 -- 810 & (5)\\
\hspace{0.1in} PACS & 53 -- 260 & 9.4$''$ & 1000 -- 4000 & 75 -- 300 & (6) \\
\hspace{0.1in} HIFI & 157 -- 625 & 10$'' - 40''$ & $> 5 \times 10^5$ & 0.15 -- 0.6 & (7) \\\hline
\end{tabular}
\begin{tablenotes}[para]
\footnotesize
\item[1] ISO, operated primarily by the European Space Agency (ESA) was in operation from  1995 to 1998.   
SWAS was a National Aeronautics and Space Administration (NASA) mission that observed from 1998 -- 2004, while Odin is primarily operated by the Swedish Space Corporation and was launched in 2001.    The \spitz\ {\em Space Telescope} (NASA) performed spectroscopy during its cooled mission from 2003 through 2009.
The {\em Herschel} {\em Space Observatory} is an ESA satellite with NASA participation that was launched in 2009 and is expected to operate for $\sim$3.5 years. \\
\item[2] (1) \citet{degraauw96}; (2) \citet{melnick_swas}; (3) \citet{odin};  (4) \citet{houck04}; (5) \citet{griffin10}; (6) \citet{poglitsch10}; (7) \citet{degraauw10}\\
\item[3] Odin also had the capability to observe at 118.75 GHz.\\
\end{tablenotes}
\end{threeparttable}
\end{table}

\subsection{Chemical Perspective}
\label{sec:chem}
 
 Oxygen is the third most abundant element in the Universe and its gas-phase chemistry is linked to products of cosmic ray ionization of \Htwo\ and the eventual formation of the trihydrogen ion
(\Hthreep ).  \Hthreep\ will react with O and a sequence of rapid
reactions produces \HthreeOp . 
The next step involves the dissociative recombination of \HthreeOp\ 
with electrons:

\begin{eqnarray}
\rm{H_3O^+ + e^-}  &\rightarrow & \rm{H + H_2O}\;\;\;\;\;\;\;\;\;\;\: f_1\\
& \rightarrow & \rm{OH + H_2}\;\;\;\;\;\;\;\;\;\;\:  f_2 \nonumber\\
&\rightarrow & \rm{OH + 2H}\;\;\;\;\;\;\;\;\;\;\: f_3 \nonumber\\
&\rightarrow & \rm{O + H + H_2}\;\;\;\;\;\;\:f_4\nonumber
\end{eqnarray} 

\noindent where $f_{1-4}$ are the branching ratios.\footnote{
Two measurements using a storage ring give roughly consistent results
\citep{jensen_h3op, vc_h3op}.   We provide here the latest measurement:
$f_1 = 0.25$, $f_2 = 0.14$, $f_3 = 0.60$, and
$f_4 = 0.01$.   Using a different technique (flowing afterglow), \citet{williams_h3op} find 
$f_1 = 0.05$, $f_2 = 0.36$, $f_3 = 0.29$, $f_4 = 0.30$.  }  Via this sequence of reactions water vapor is then created.    The primary {\em gas-phase} destruction pathway in shielded gas is via a reaction with He$^+$ (another ionization product); at the disk surface photodissociation also plays a key role.   These reactions for pure gas phase chemistry result in an abundance (relative to H$_2$) of a few $\times 10^{-6}$.   It should be stated that the example above provides only the primary formation and destruction pathways under typical conditions (i.e. where ionizing agents are available).  In addition, water participates in numerous other reactions both as reactant and product and some water can be created via other pathways, albeit at reduced levels \citep[e.g.][]{bergin_impact}.

A critical factor in cold gas is the likelihood that water is created in situ on grain surfaces via  a series of reactions starting with atomic O and H and forming water ice \citep{th82, cuppen10}.    In fact water ice is observed with abundances close to $\sim 10^{-4}$ (relative to H$_2$) along several lines of sight \citep{gibb_iso, sonnentrucker08}.
The evaporation temperature of water ice at interstellar pressures is $\sim$ 100 K \citep{fraser_h2obind}.   Thus water will likely remain frozen for any evolutionary stage characterized by temperatures below 100 K.
This can have a direct impact on the abundance of water vapor as the fuel for the gas-phase chemistry is oxygen atoms.  When oxygen atoms are mostly trapped on grain surfaces as water ice the gas-phase chemistry is incapable of maintaining abundances of $\sim 10^{-6}$ and the predicted abundance is much lower \citep{bergin_impact, cre01, rh02}.      Since a large fraction of the mass in star-forming cores will be below the sublimation temperature, it has been proposed that non-thermal desorption mechanisms may be needed to release water from the ice mantle coating the grain surface.  In particular, ultraviolet photodesorption is found to be efficient in laboratory experiments \citep{westley_h2opd, olvv10}.   Given the prevalence of UV photons above the Lyman limit in the vicinity of star-forming regions and protoplanetary disks, it is possible that this mechanism would play an important role \citep{dchk05, hkbm09}.

At temperatures below a few hundred Kelvins  ion-neutral chemistry dominates.  Above this temperature, two neutral-neutral reactions rapidly transform all elemental oxygen in the molecular gas into water vapor
\citep{edj78, wg87}:

\begin{eqnarray}
\rm{O  +   H_2} &   \rightarrow & \rm{  OH   +   H}\;(E_a= 3160\;\rm{K}),\\
\rm{OH + H_2} & \rightarrow & \rm{H_2O + H}\;(E_a= 1660\;\rm{K}).
\end{eqnarray}

\noindent This mechanism (along with evaporation of water ice) will keep water vapor the dominant oxygen component when the gas temperature exceeds $\sim$300 K \citep{kn96a}.

\section{H$_2$O in Star and Planet Formation}

Prior to \her\ the observational perspective of water vapor pointed to some intriguing issues with respect both to its disposition as solid or vapor and also its destruction.   ISO found clear evidence for hot water associated with shocks but also perhaps isolated in the hot cores of both low and high mass stars \citep[see the review of][]{evd04, nisini05}.   SWAS and Odin discovered that cold water vapor is less abundant than expected from pure gas-phase chemistry, hinting at a role of freeze-out and ice formation \citep{snell_swash2o, hjalmarson03}.    \spitz\ was readily suited to search for water emission in young protoplanetary disks \citep{carrnajita08}, while all of the above was supplemented by ground-based observations using masing transitions in open atmospheric windows or by observing  isotopologues which have lower abundance.
In the following we attempt to note the important contributions from the various space-based platforms.   However, in general we will highlight more recent results.   Thus we will focus in large parts on the advantages of \her\ to probe hot, warm, and cold water, but also the ability to spectrally resolve the line profiles with sensitivity.  In the following we will trace our knowledge of water from the large parsec-sized cloud scale  down to an astronomical unit within the terrestrial planet-forming regions of circumstellar disks.

\subsection{Cloud}

Molecular clouds are extended on large, degree-sized, spatial scales which makes it difficult to probe the full extent of water emission.
The largest water maps that are in existence are the maps of Orion obtained using weak masers observed from the ground \citep{cerni94, cerni99}, and of the o-\HtwoO\ ground state by SWAS \citep{melnick11}, and Odin \citep{olofsson_odinbnkl}.  An additional map was made of M17 \citep{snell_m17}.
These maps demonstrate that water emission is present on  $\sim 1$~pc size scales.
However, this does not correspond to the cloud as traced by $^{12}$CO, but rather on scales more representative of the denser star-forming cores traced by C$^{18}$O.    Based on analysis of these data the average water abundance is found to be rather low \cite[$\sim 10^{-8}$;][]{snell_swash2o} and consistent with models that include the formation of water ice mantles and the freeze-out of water \citep{bergin_impact, cre01, rh02}.    Analysis of the masing transitions hinted at higher abundances \citep{cerni94, cerni99}, suggesting that there might be an abundance gradient along the line of sight.

Detailed comparisons of the water emission distribution and other gas-phase tracers suggest that water is not tracing the volume of the cloud, but rather is present mainly on the surface \citep{melnick11}.    Since the dust temperatures on the surface are still below that of the sublimation temperature of water ice this has brought the focus onto the potential impact of non-thermal methods to desorb the frozen water molecules.   Thus, recent modeling efforts have focused on whether photodesorption might be responsible for the observed water vapor.    Such models have been found to be consistent with earlier SWAS results \citep{hkbm09}.   Thus water vapor is photodissociated on the exposed surface of the core, but as the radiation field decays deeper in the cloud water vapor is found within a layer where photodesorption of ice is balanced by photodissociation of gas.  Deeper into the cold (T $\sim 10 - 20$ K) core the water is predominantly found as ice.   It is this type of model that will be directly tested by additional \her\ observations that hint at the link between grain physics and the chemistry.

\begin{landscape}
\begin{figure}
\centering
\vspace{-5mm}
\includegraphics[scale=0.95]{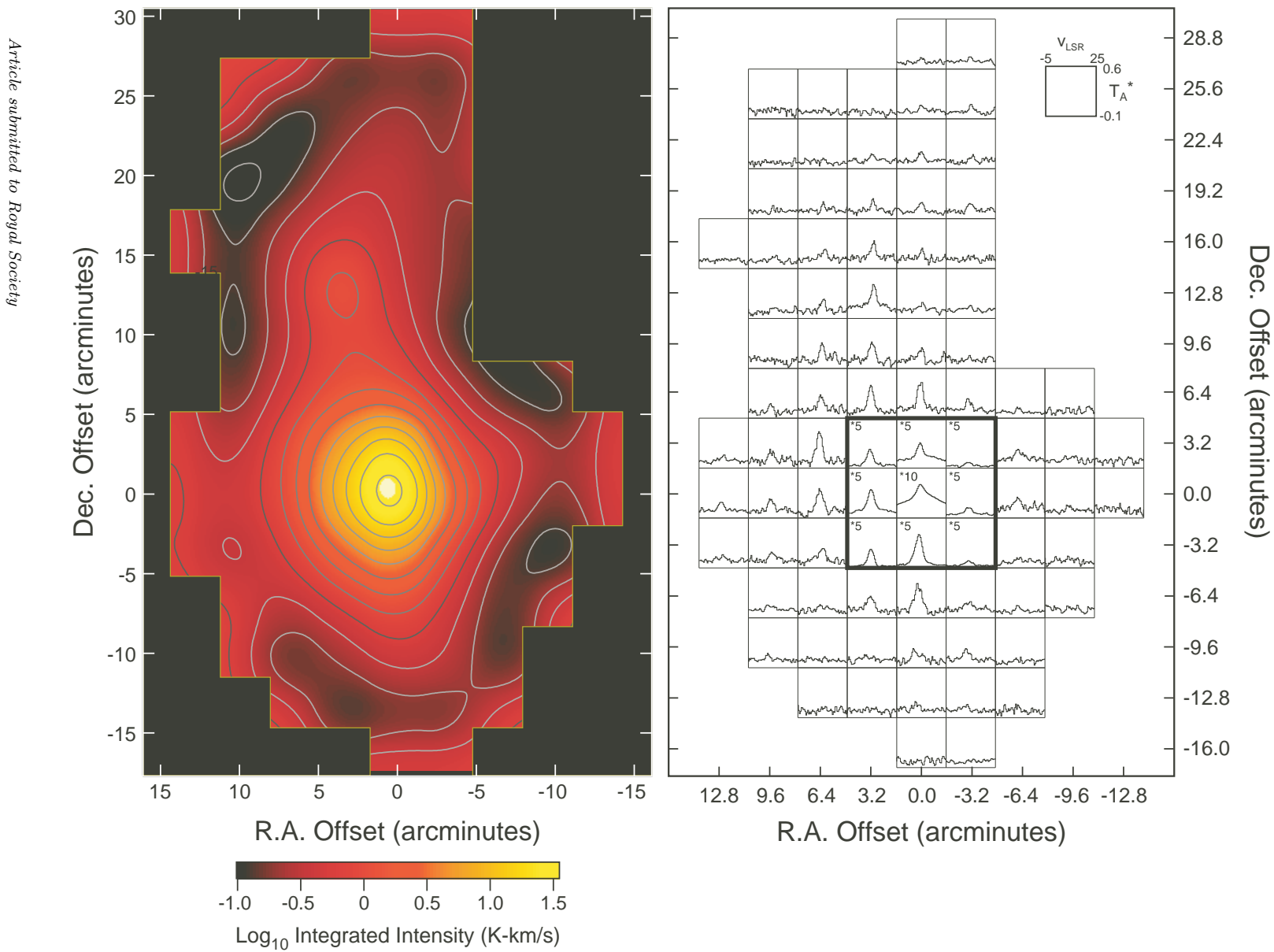}
%\includegraphics[angle=90,scale=0.76]{fig4.pdf}
%\includegraphics[angle=90,scale=0.137]{Orion_H2O_Map.pdf}
%\vspace{0.1mm}
\renewcommand{\baselinestretch}{0.97}
\caption{Integrated intensity map of the Orion Molecular Ridge in 1$_{10}$-1$_{01}$ 
556.9~GHz transition using {\em SWAS} and taken from \citet{melnick11}.   The starred numbers
within the spectra in the inner square indicate the values by which the antenna temperatures  have been divided 
so that the scaled spectra fit within this plot.
}
\renewcommand{\baselinestretch}{1.0}
\label{fig:map3}
\end{figure}
\end{landscape}

\subsection{Pre-stellar Cores}

Lacking the presence of a central stellar source which releases radiative and kinetic energy, low mass pre-stellar dense cores provide excellent laboratories to test the basic gas/solid state physics and chemistry  prior to stellar birth. 
This is strengthened by the fact that the physical structure (density and temperature) are fairly well constrained from observations of dust thermal continuum emission  \citep{andre_ppiv, shirley02}, which readily allows for exploration of molecular abundances.
 Since temperatures are $\sim 10$~K, there is general expectation that freeze-out will dominate the chemistry.  Indeed these sources exhibit wide-spread evidence for gas-phase abundance depletions of species less volatile than water, such as CO that confirms this interpretation \citep[see][and references therein]{bt_araa, difran_ppv}.     Previous observations by SWAS and Odin had set very low abundance limits in agreement with the dominance of freeze-out \citep{bergin_b68h2o, klotz08}, and a large amount of water present as ice \citep{whittet10}.

\begin{figure} 
\centering
%\subfigure[Theory]{
\includegraphics[scale=0.7]{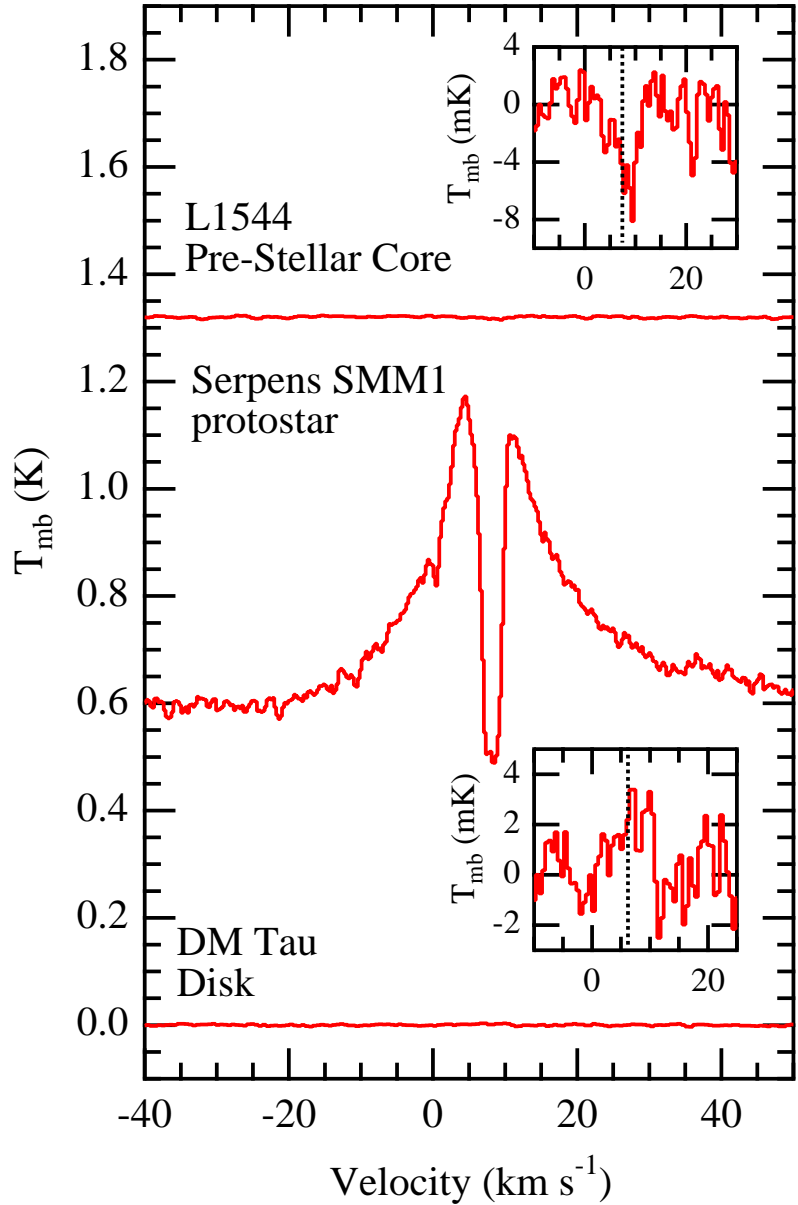}
%\label{fig:subfig1}}
%\subfigure[Observations]{
%\includegraphics[scale=0.5]{bergin-dmtau-fig2.eps}
%\label{fig:subfig2}}
\caption{1$_{10}$ -- 1$_{01}$ o-H$_2$O emission spectra obtained with {\em Herschel}/HIFI spanning evolutionary stages that trace the formation of low-mass stars and planets.   The pre-stellar core L1544 is shown on top, along with an expanded view of the central regions of a baseline subtracted spectrum near the system velocity.   The middle spectrum illustrates the strong and broad emission seen toward the Serpens SMM1 protostar.    The bottom spectrum shows the tentative detection towards the DM Tau protoplanetary disk along with an expanded view near the systemic velocity.
In the case of the blow-up region of L1544 and DM Tau the systemic velocity is given as a dashed line.
Each of these spectra has been presented and analyzed previously \citep[][]{caselli10, bergin10a}. }
\label{fig:lm}
\end{figure}

Initial surveys by \her\ focused on detecting the ground state line of o-\HtwoO\ in two template objects B68 and L1544 \citep{caselli10}.  In the case of B68 no emission was found with a 3$\sigma$ limit on the abundance $x$(o-\HtwoO ) $< 1.3 \times 10^{-9}$ (relative to \Htwo ).   In Fig.~\ref{fig:lm} we show the spectrum of L1544 along with objects at different evolutionary stages.   As seen in the expanded view of the spectrum coincident with the source velocity there is a surprising detection of water in absorption.  This absorption is seen at the 5$\sigma$ level and represents the first detection of water vapor in a cold (T $\sim 10$ K) object. 
Initial analysis suggests that the water abundance peaks away from 0.1 pc the core center with an abundance $\sim 10^{-8}$, again with water inferred to be present mostly in the form of ice \citep{caselli10}.   A deeper observation of this line towards L1544 is planned to confirm this result.  

\subsection{Protostars and Outflows}

\subsubsection{Low-Mass Star Forming Regions}

Low-mass solar type protostars are important as tracing the stage where the star builds up its mass and the disk forms from the collapsing envelope.   An important aspect of water in these regions is the expected large abundance ($\gtrsim 10^{-4}$, relative to \Htwo ) of water vapor that might result from evaporating ices or high temperature reactions.  This large abundance would lead water to be an important coolant both in the inner envelope close to the star \citep{cht96} and also in the outflow \citep{kn96a}.  Thus water can participate in the physics of collapse and mass ejection.  Initial studies by ISO found evidence for strong water emission and high abundances with water playing an important role as a gas coolant \citep{ceccarelli99, giannini01}.   However, the exact disposition of the water was in doubt.   One theory posited the water as tracing molecular outflows \citep{molinari00, ngl02}, other theories favored the inner collapsing envelope which is warm enough to evaporate ices \citep{ceccarelli99, maret02}, or perhaps due to envelope accretion shocks \citep{watson07}.   High spectral resolution observations by SWAS and Odin found evidence for broad line wings in the spectrum of \oHtwoO\ $1_{10} - 1_{01}$ towards representative protostars \citep{franklin08, bj09}.  This is clearly associated with the outflow, but the low spatial resolution hampered any  search for emission associated with the protostar. 

A \her\ survey of water emission towards protostars reveals that water emission is quite common and dominated by outflows/shocks \citep[][and Kristensen et al. 2011, in prep.]{kristensen10, kvd11}.    One sample spectrum of a low mass protostar as seen by \her\ is given in Fig.~\ref{fig:lm} where we show the o-\HtwoO\ $1_{10} - 1_{01}$ transition observed towards Serpens SMM1.    This spectrum shows representative features -- a broad line profile with full width at half-maximum of $> 20$ km s$^{-1}$ and evidence for a narrow absorption feature at the systemic velocity, both with line widths (at half maximum) of a $\sim$few km s$^{-1}$.   A survey of over 29 objects in the ground state line of o-\HtwoO\ finds similarly complex profiles, with broad emission lines and narrow absorption/emission.  Higher excitation lines are present, but also are found to have broad ($>$ 10 km s$^{-1}$) line widths \citep[][and Kristensen et al. 2011, in prep.]{kristensen10}.   Even the lines of \HtwoeiO , a factor of $\sim 500$ less abundant, are still broad.

In all we have learned that water emission in low mass sources shows a clear dominance of emission from the outflowing gas.
This emission is attributed to shocks in the inner envelope and analysis suggests that the abundance ratio of H$_2$O/CO increases with velocity with water trapping $\sim$5-10\% of the available oxygen at the highest velocities \citep{kristensen10}.   The narrow emission/absorption components from the envelope are consistent with abundances of $\sim 10^{-8}$ (relative to H$_2$).  This is comparable to that seen in pre-stellar cores and again points to the importance of water ice and freeze-out.   Deep searches for high-J H$_2^{18}$O have begun to reveal water in the inner envelope but abundances ($\sim 10^{-5}$, relative to \Htwo ) appear to be lower than the expected value of 10$^{-4}$  (Visser et al. 2011, in prep.).  
 
 \begin{figure} 
\centering
%\subfigure[Theory]{
\includegraphics[scale=1.0]{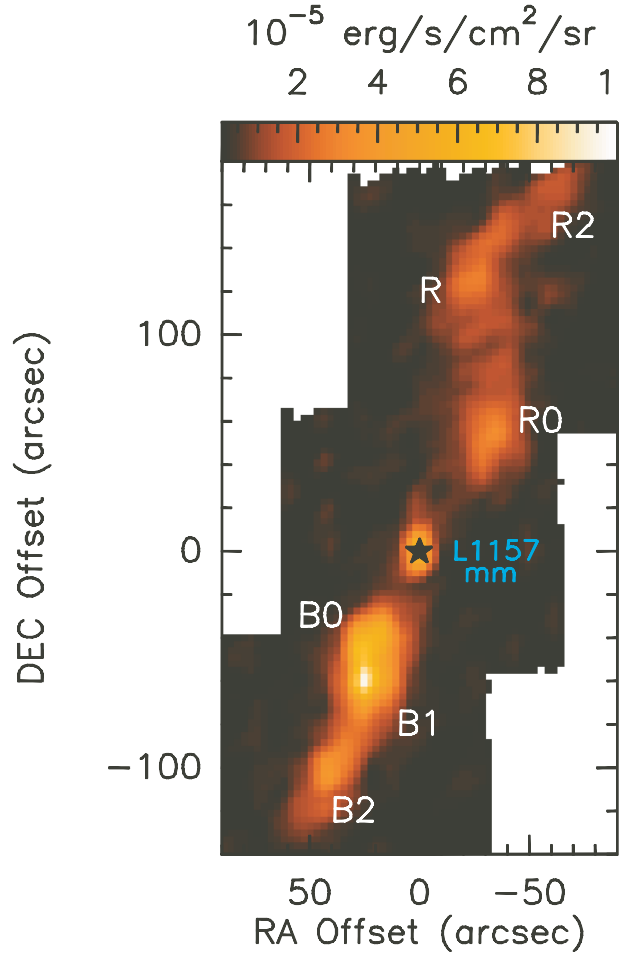}
%\label{fig:subfig1}}
%\subfigure[Observations]{
%\includegraphics[scale=0.5]{bergin-dmtau-fig2.eps}
%\label{fig:subfig2}}
\caption{Map of the continuum-subtracted H$_2$O 179 $\mu$m emission in the L1157 outflow published by \citet{nisini10}.
The mm-continuum source coincident with the driving source is noted with a star; other peaks within the blue-shifted lobe (B) and red-shifted lobe (R) are also noted.
}
\label{fig:l1157}
\end{figure}

The picture of water in low mass stars is not complete without looking beyond the central pixel associated with the protostar.   Low mass sources are well known to have extended bipolar outflows that have been traced in a variety of species, including H$_2$O \citep{bachiller96}.    The direct association of water with molecular outflows is shown in the beautiful emission map of the 179 $\mu$m line of o-\HtwoO\ seen towards L1157 \citep{nisini10}, which is given in Fig.~\ref{fig:l1157}.    Analysis of these data shows that water is closely associated with hot \Htwo\  emission associated with the flow \citep{nisini10a}, and with other shock tracers such as SiO.     The \Htwo\ emission traces shocked gas with temperatures of $\sim$300 -- 1300~K, which is clearly capable of powering the neutral-neutral reactions discussed in \S~3 or perhaps water is formed via sputtering of the ices \citep{jth96}.  Regardless, for this source, water is formed in abundance ($x$(\Htwo) $\sim$  10$^{-5}$ -- 10$^{-4}$) and is a major coolant in the {\em high} velocity gas \citep{nisini10, lefloch10}.  Whether this is representative of all molecular outflows is an outstanding issue and will be answered by additional \her\ WISH observations.
 \begin{figure}
\includegraphics[width=8cm, angle=-90]{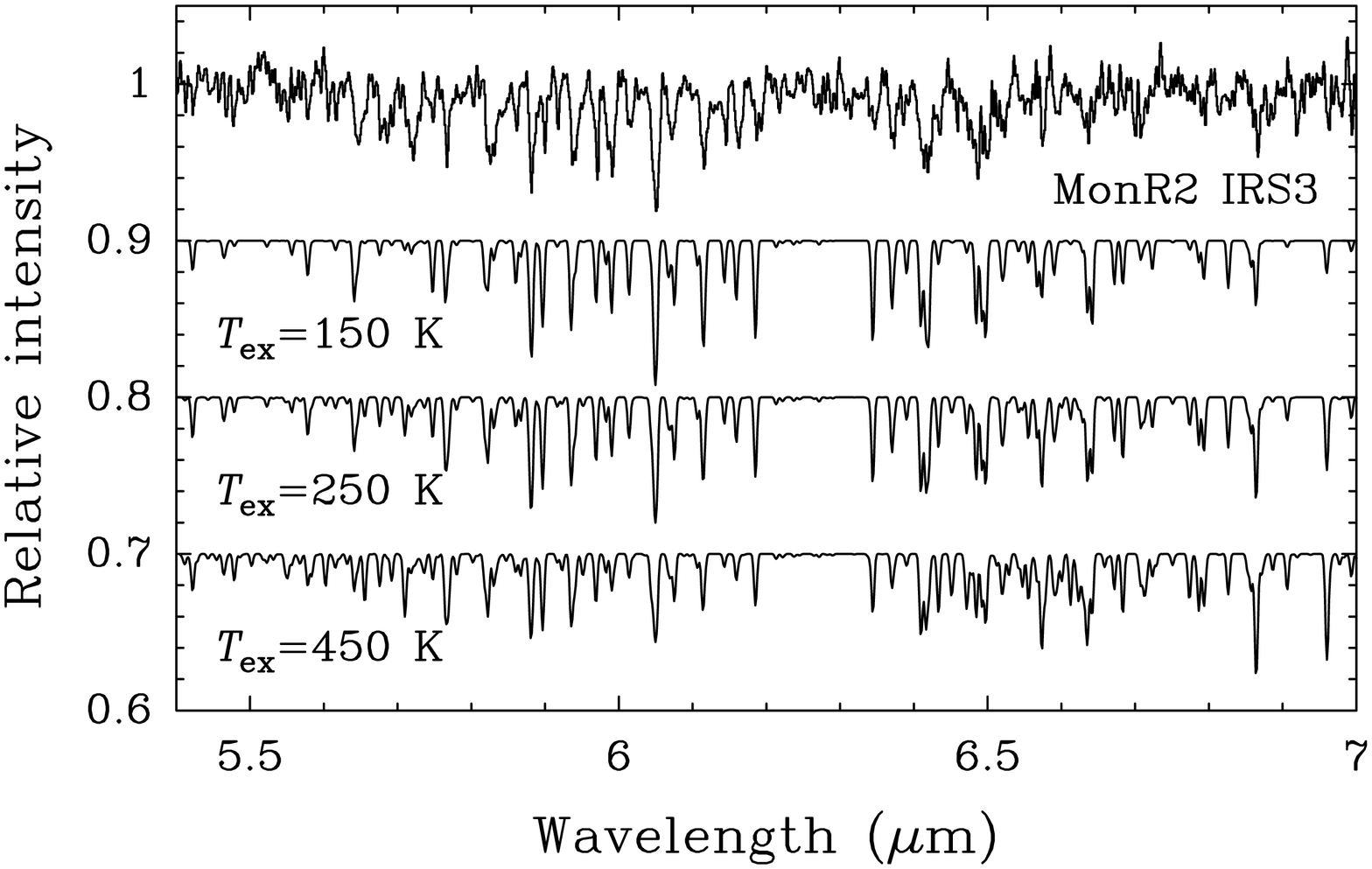}
\caption{ISO-SWS observations of the ro-vibrational $\nu_2$ band of gas-phase H$_2$O
toward the high-mass protostar MonR2 IRS3 at R = 1500. The data are compared with
model spectra at different excitation temperatures. These absorption
data probe the amount of water in the innermost part of the envelope
whereas emission data are more sensitive to the outer part \citep[from:][]{bvd03}.
\label{fig:iso}}
\end{figure}

\subsubsection{High-Mass Star Forming Regions}

Stars significantly more massive than our Sun are important in astronomy as they dominate the life cycles of galaxies.
Moreover, young massive protostars are surrounded by a rich envelope of molecular material that are the chemically richest sources in the Milky Way \citep{hvd09}.   In fact it has long been known, based on ground-based observations of H$_2^{18}$O that massive hot cores harbor large amounts of water vapor \citep{jacq88, gmw96, vdt06}, this was further supported by ISO observations \citep{helmich96a, cerni_isobnkl, boonman03, lerate10}.   A sample of the rich ISO data of high mass stars is shown in Fig.~\ref{fig:iso}.
Here the 6 $\mu$m $\nu_2$ band of water is observed in absorption towards Mon R2 IRS3.   A comparison of model and data show that the gas in the inner envelope is clearly hot and rich in water vapor with abundances $\sim 10^{-5}$ \citep{bvd03}.  One of the strengths of ISO was the ability to observe both the gas (shown here) and the solids \citep[e.g.][]{boogert00, gibb_iso} and in high mass sources the gas/solid ratio increases with temperature hinting at some contribution from grain mantle evaporation \citep{bvd03}.       

\begin{figure}
\hspace{-0.2in}
\subfigure[Orion KL]{
\includegraphics[scale=0.3]{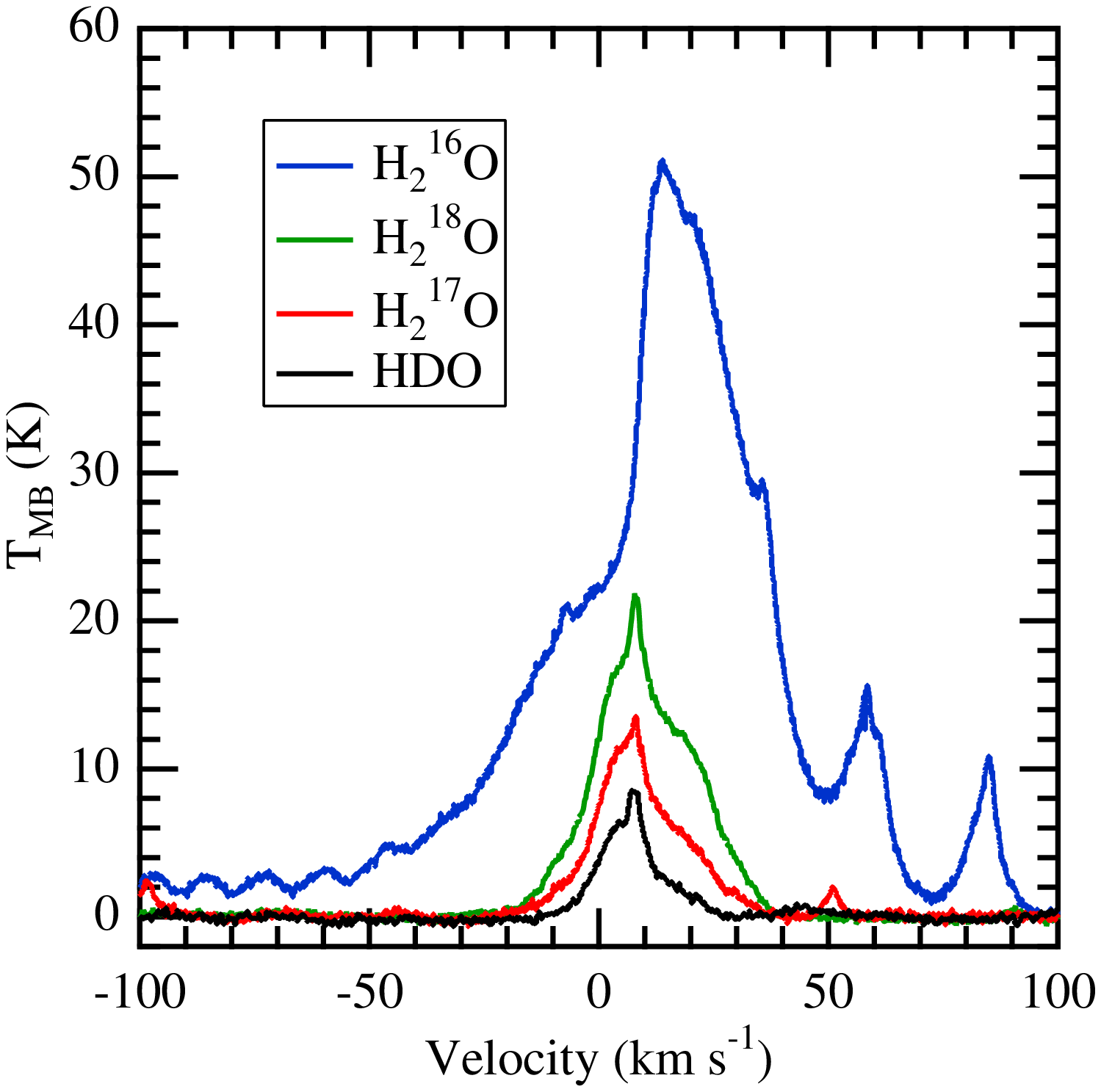}
\label{fig:specsub1}}
\subfigure[W3 (IRS5)]{
\includegraphics[scale=0.6]{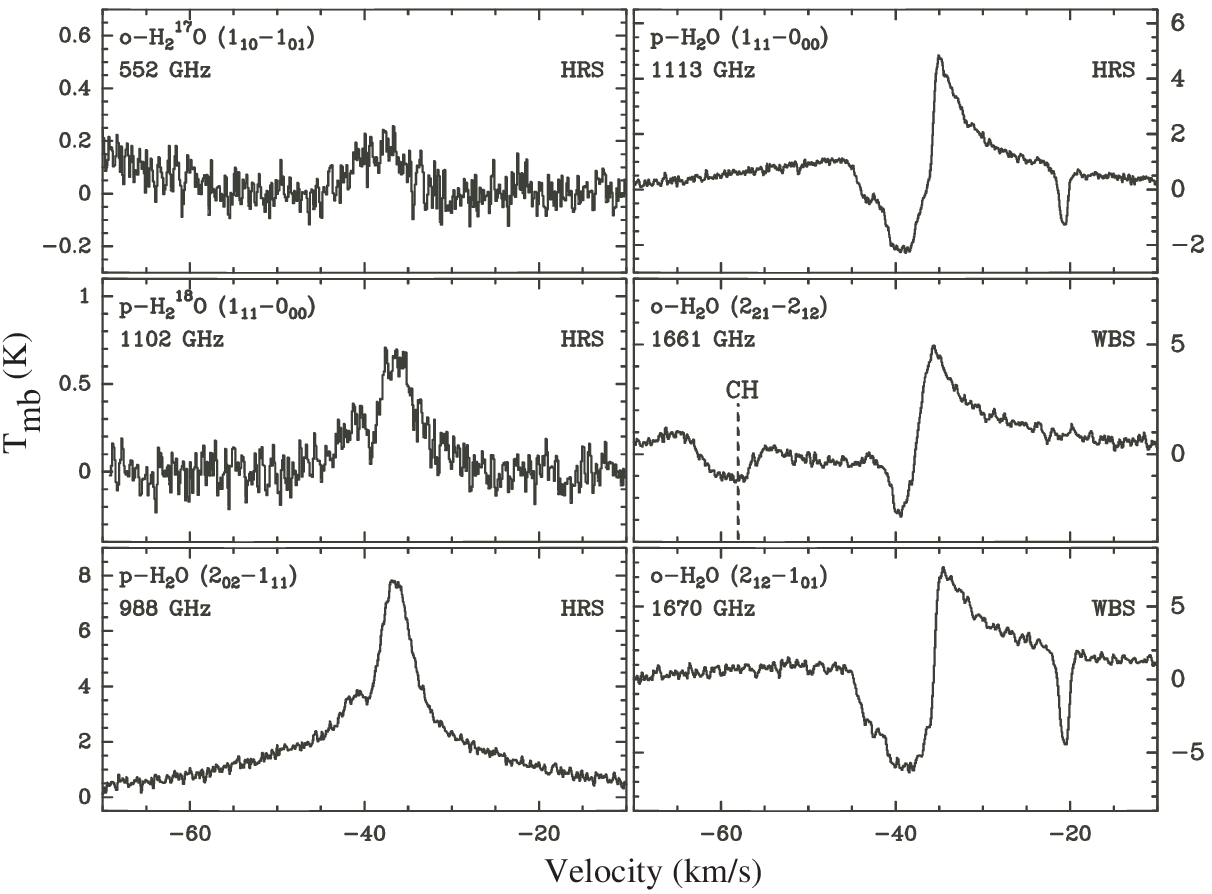}
\label{fig:specsub2}}
\caption{\subref{fig:specsub1} Sample of the $3_{21} - 3_{12}$ emission lines of water and its isotopologues around 1 THz detected towards Orion KL \citep{melnick10}. \subref{fig:specsub2} Selection of water lines observed towards W3 IRS5 from \citet{chav10}.
}
\label{fig:hm}
\end{figure}

Fig.~\ref{fig:hm} presents \her\ observations of Orion KL and W3 IRS 5 , which are representative of massive star-forming cores.  This illustrates several facets of these regions.   First because of the high temperatures, high densities (\nhtwo\ $> 10^6$ \cc ), and large gas columns ($> 10^{23}$ cm$^{-2}$) numerous water lines can be detected.   In Orion the lines are strong enough to observe a suite of isotopologues with over 50 detected transitions \citep{melnick10}.  This is shown in Fig.~\ref{fig:hm}~\subref{fig:specsub1} with detections of the  $3_{21} - 3_{12}$ transition of H$_2^{16}$O, H$_2^{18}$O, H$_2^{17}$O, and HDO.  This is even extended to a detection of HD$^{18}$O \citep{bergin10a}.    Second, the lines profiles are very complex with different transitions having significantly different structure.  In W3 IRS5 there is evidence for cold foreground absorbing spectral components (often more than one) superposed on broad ($> 5 - 10$ km/s) components arising from outflows and also from a zone within the envelope where ices are evaporated  \citep{chav10}.    
 
Models of the line emission utilize the standard procedure of estimating the physical structure from the overall dust emission and its spectral energy distribution \citep{vdt00, marseille08}.    Within the envelope the water abundance assumes a jump structure.  When the dust temperature exceeds the sublimation temperature of water ice of $\sim 100$ K the abundance is elevated; this traces the hot core.   In the colder outer envelope the abundance is another variable, but is always found to be below that of the hot core.    With these assumptions some clear facts are emerging \citep{vdt10, chav10, empr10, marseille10}.  (1)  The abundance of hot water in the inner hot core is elevated to high values $> 10^{-5} - 10^{-4}$ (relative to H$_2$).   (2) The outer envelope has water vapor abundances $\sim 10^{-8}$, which is below that expected from pure gas-phase chemistry alone.    This is interpreted to be the influence of the freeze-out of water onto the surfaces of dust grains.

Since massive star-forming regions are more distant even the improved angular resolution of \her\ does not always provide a resolved picture of the bipolar outflows.  Moreover there are often more than one source within the telescope beam which further complicates the analysis.  However,  in general the molecular outflows are clearly associated with water emission as is evidenced by the broad line wings shown in Fig.~\ref{fig:hm}.    Detailed analysis of the Orion spectrum finds an abundance of $\sim 10^{-4}$ \citep{lerate10, melnick10}, which confirms the important role that water plays as a major coolant in shocked molecular gas \citep[e.g][]{kn96a}.

 \begin{figure}[t]
%\epsscale{1.5}
%\centerline{\psfig{figure=tdep-tevap.eps,height=5in}}
%\includegraphics[width=9cm, angle=-90]{Carr_SciFig1.eps}
\includegraphics[width=9cm, angle=-90]{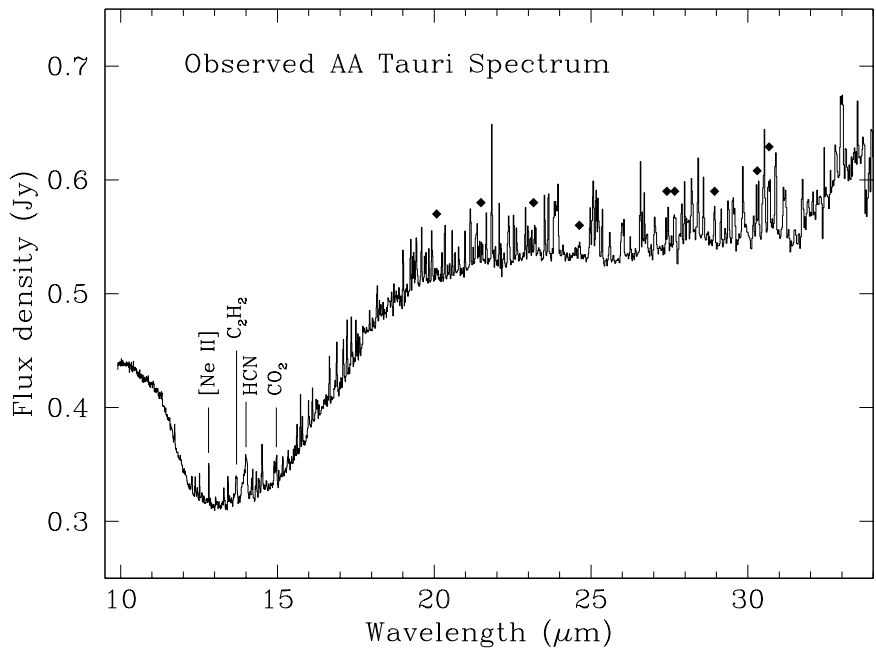}
\caption{Infrared spectrum of the classical T Tauri star AA Tau taken by the {\em Spitzer Space
Telescope}.  Vibrational modes of C$_2$H$_2$, HCN, and CO$_2$ along with a transition of  [Ne II] are labelled.
  Diamonds mark detections OH rotational transitions.  Numerous rotational transitions of water vapor (not marked) 
  are spread throughout the spectrum.  Figure and discussion published by\citet{carrnajita08}.
\label{fig:cn}}
\end{figure}

\subsection{Protoplanetary Disks}

What we have learned from our study of dense cores is that water is provided to the planet-forming disk primarily as ice.
Within the disk itself there is the potential for complex motions that could alter the water ice reservoir
that originated from the parent cloud \citep{bergin_ppv, semenov10}.   However, there is a general expectation for water vapor to exist inside the radius where the temperatures rise above the sublimation temperature.  This is the so-called
``snow-line''.   Inside this radius water is found in the gas and beyond the snow-line water resides predominately as ice \citep{sasselov00, min_snowline11}.     Despite the observational challenge of observing systems with small angular size
($<$ 1-2$''$), the distribution of water vapor in protoplanetary disks is now becoming clear.

\begin{figure} 
\centering
%\subfigure[Theory]{
\includegraphics[scale=0.45]{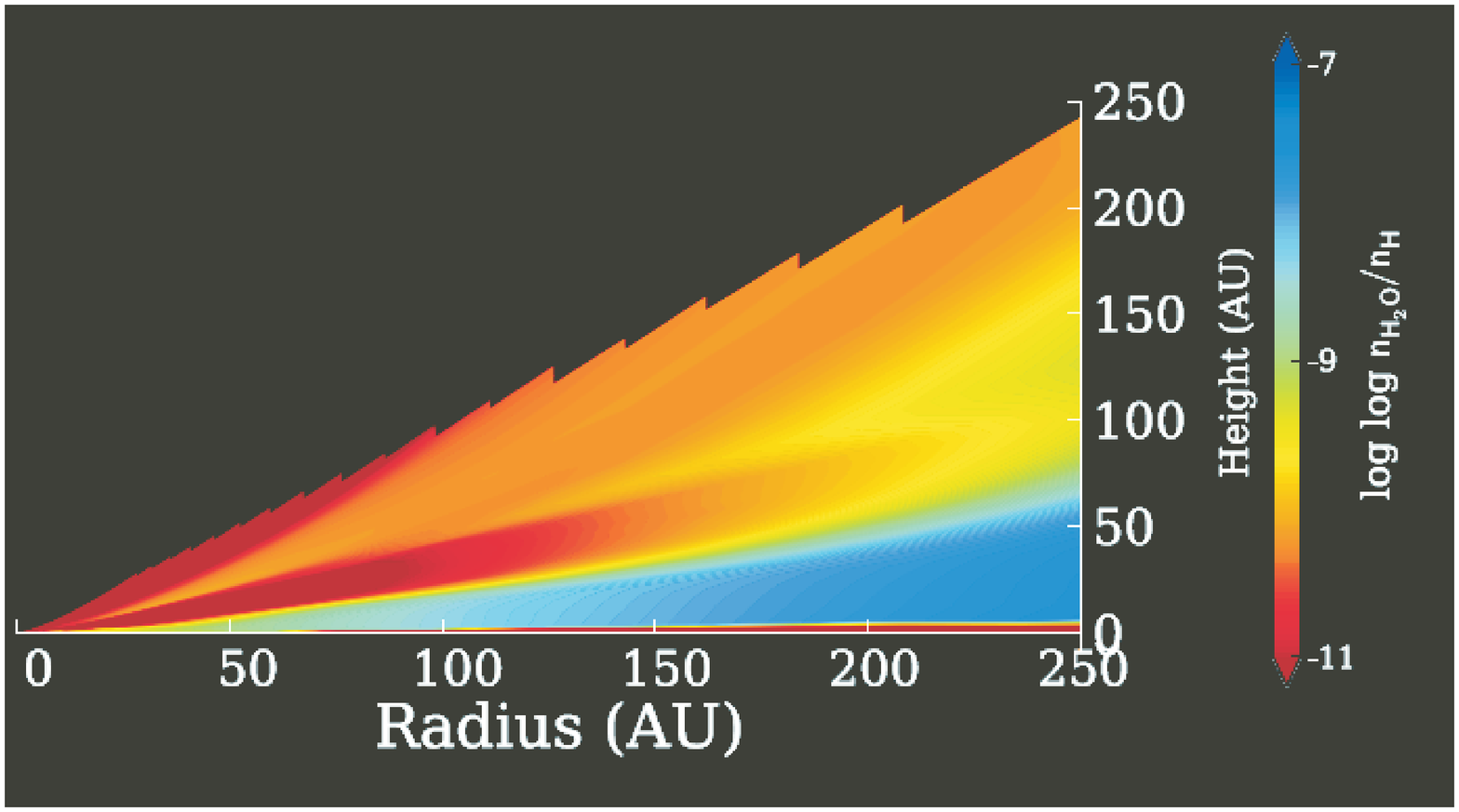}
%\label{fig:subfig1}}
%\subfigure[Observations]{
%\includegraphics[scale=0.5]{bergin-dmtau-fig2.eps}
%\label{fig:subfig2}}
\caption{Theoretical model of disk water chemistry showing the water vapor abundance relative to total H as a function of disk radial and vertical height..   This model presented uses the results given by \citet{bergin10b} and \citet{fogel11}.}   %\subref{fig:subfig2} present an the observed spectra of the ground state lines of ortho- and para-H$_2$O towards the DM Tau disk.  Results given in \citet{bergin10b}.}
\label{fig:dmtau}
\end{figure}

Surveys of water vapor emission at near to mid-infrared wavelengths using \spitz\ ground-based telescopes have revealed that water is common around Sun-like stars \citep{carrnajita08, salyk08, pont10}.   
A sample of the rich spectrum of emission lines seen towards the AA Tau protoplanetary disk is given in Fig.~\ref{fig:cn}.
These studies have shown that the  emission arises from within 0.1-10 AU tracing gas temperatures of $\sim 500 - 1000$~K and with abundances that approach that of carbon monoxide.      Because the disk surface is heated by stellar irradiation it is likely that the emission arises from warm surface layers where water is produced {\it in situ} by the high temperature reactions discussed in \S~\ref{sec:ma} \citep{gmn09, bb09}.  
Models of the emission suggest it is truncated at radii in excess of 1 AU and suggestive of a lack of water vapor in the upper layers at larger radii \citep{meijerink09}.      As demonstrated in Fig.~\ref{fig:h2ospec}, the \spitz\ wavelength range is optimized towards the detection of warm/hot water (T $> 200$~K) emission lines that will probe the terrestrial planet forming region.
\her\ observations will sample colder gas at larger radii and eventually will delineate the location of the water snow-line in a variety of systems.
It is also worth noting that water vapor emission is generally lacking in the inner disks near young massive stars \citep{mandell08, pont10, fedele11}.

Beyond the snow-line the disk temperature is below the sublimation point for water ice.    However, theory predicts that photodesorption can provide a tenuous layer of water vapor on the disk surface \citep{dchk05}.   Since UV photons are plentiful in young stars that are still accreting from the circumstellar disk \citep{bergin_lyalp}, this provides for the possibility to find water at large distances from the star.   Results from  a detailed model of the chemistry are shown in Fig.~\ref{fig:dmtau}.   This model shows the layer of photodesorbed water exists above the water-ice rich (water vapor depleted) midplane.     Given that the gas temperatures are below 100 K the strongest emission lines lie in the far-infrared wavelengths (e.g. Fig.~\ref{fig:h2ospec}.   A search for this emission was reported by \citet{bergin10b}.  A very tentative detection was reported in the o-H$_2$O (1$_{10}$ -- 1$_{01}$) ground-state transition, which is shown in Fig.~\ref{fig:lm}.  The interesting aspect from these observation was that the models shown in  Fig.~\ref{fig:dmtau} predict emission lines in excess of the observational limits.   A possible answer to this issue is if the upper layers that are exposed to UV have, over time, become depleted in the water-ice coated grains.  A detection of water vapor in this cold layer will be reported by Hogerheijde et al. 2011 (in prep.).
Regardless these observations support the picture of the outer disks being dominated by ice.

\section{Summary}
\label{sec:sum}

 This high level overview overlooks many of the numerous details or unknowns that are still present.  
 For example we are only now grappling with the full distribution of water along the line of sight and what types
 of shocks (and where) the water is forming.    Moreover, the role of radiation is clearly important as a mechanism to release frozen water via
 photodesorption, but models need to be more directly confronted with observations in a range of environments.   Radiation is also important for the destruction of water and limiting its abundance, but peering close to heavily embedded objects or in the interiors of protoplanetary systems is difficult.   More observations and theory will be needed to tease out the exact ways water forms and is destroyed.   Nonetheless  it is fair to state that over the past 20 years we have undergone a revolution of our understanding of water in star forming regions and portions of the water cycle have been revealed.   
The early evolutionary stages are dominated by the formation of a water ice mantle that traps a significant ($\gtrsim
 10$\%) amount of the available oxygen.    Thus the gas to solid ratio prior to stellar birth is low,  H$_2$O$_{gas}$/H$_2$O$_{ice}$ $\sim 10^{-4} - 10^{-3}$.  When a star is born several things happen (1) the water in the inner envelope evaporates, but may be limited by stellar radiation.  (2)   Water is clearly formed and associated with gas that is in motion.   Put another way, water is associated with shocks in the outflowing gas - both close to the star and in the bipolar flow at greater distances.  This dominates the water vapor emission profile.   (3)  Water is provided to the planet forming disk predominantly as ice.   
   Within the disk itself water is clearly present and a snow-line or evaporation front on the disk surface is inferred to be present, but the exact location of the snow-line in the disk midplane is yet to be constrained.   Ice again dominates the disk mass and it is possible that it originated in the pre-stellar core before the star was born.

\acknowledgements

This work summarizes the current state-of-the-art regarding our knowledge of water beyond the solar system.  This summary would not
have been possible without the fantastic work of the instrument builders of the various space missions outlined above and we are 
exceedingly grateful to their efforts.   Similarly we are grateful to chemical physicists for providing a wide array of grounding
data (collision rates, frequencies, line strengths, solid state properties, etc.) that is needed for proper interpretation.  We are also 
grateful to excellent work of the \her\ HEXOS, WISH, and CHESS teams, whose initial results are given above.
The authors thank L. Kristensen for providing reduced \her\ data
as needed for display.  The work of EAB is supported by funding from NASA through an award issued by JPL/Caltech (Herschel GTO).  
EvD is supported by NOVA, by a Spinoza grant
and grant 614.001.008 from NWO, and by EU FP7 grant 238258.

\newpage 

\newcommand{\nat}{{\em Nature }}
\newcommand{\aap}{{\em Astron. \& Astrophys. }}
\newcommand{\aj}{{\em Astron.~J. }}
\newcommand{\apj}{{\em Astrophys.~J. }}
\newcommand{\araa}{{\em Ann. Rev. Astron. Astrophys. }}
\newcommand{\apjl}{{\em Astrophys.~J.~Letters }}
\newcommand{\apjs}{{\em Astrophys.~J.~Suppl. }}
\newcommand{\apss}{{\em Astrophys.~Space~Sci. }}
\newcommand{\icarus}{{\em Icarus }}
\newcommand{\mnras}{{\em MNRAS }}
\newcommand{\pasp}{{\em Pub. Astron. Soc. Pacific }}
\newcommand{\planss}{{\em Plan. Space Sci. }}
\newcommand{\physrep}{{\em Phys. Rep.}}
\newcommand{\bain}{{\em Bull.~Astron.~Inst.~Netherlands }}

\bibliography{/Nirgal1/ebergin/tex/bib/ted}

\end{document}